\documentclass[dvips]{acta}
\usepackage{supertabular,lscape,epsfig}
\usepackage{amssymb}
\usepackage{amsmath}
\usepackage{float}
\mathchardef\mhyphen="2D
\usepackage{placeins}
\DeclareSymbolFont{ppa}{OT1}{ppl}{m}{it}
\DeclareMathSymbol{\vv}{\mathalpha}{ppa}{'166}

\SetPages{1}{18}

\SetVol{76}{2026}

\begin{document}
\newcommand\pvalue{\mathop{p\mhyphen {\rm value}}}
\newcommand{\TabApp}[2]{\begin{center}\parbox[t]{#1}{\centerline{
  {\bf Appendix}}
  \vskip2mm
  \centerline{\small {\spaceskip 2pt plus 1pt minus 1pt T a b l e}
  \refstepcounter{table}\thetable}
  \vskip2mm
  \centerline{\footnotesize #2}}
  \vskip3mm
\end{center}}

\newcommand{\TabCapp}[2]{\begin{center}\parbox[t]{#1}{\centerline{
  \small {\spaceskip 2pt plus 1pt minus 1pt T a b l e}
  \refstepcounter{table}\thetable}
  \vskip2mm
  \centerline{\footnotesize #2}}
  \vskip3mm
\end{center}}

\newcommand{\TTabCap}[3]{\begin{center}\parbox[t]{#1}{\centerline{
  \small {\spaceskip 2pt plus 1pt minus 1pt T a b l e}
  \refstepcounter{table}\thetable}
  \vskip2mm
  \centerline{\footnotesize #2}
  \centerline{\footnotesize #3}}
  \vskip1mm
\end{center}}

\newcommand{\MakeTableApp}[4]{\begin{table}[p]\TabApp{#2}{#3}
  \begin{center} \TableFont \begin{tabular}{#1} #4
  \end{tabular}\end{center}\end{table}}

\newcommand{\MakeTableSepp}[4]{\begin{table}[p]\TabCapp{#2}{#3}
  \begin{center} \TableFont \begin{tabular}{#1} #4
  \end{tabular}\end{center}\end{table}}

\newcommand{\MakeTableee}[4]{\begin{table}[htb]\TabCapp{#2}{#3}
  \begin{center} \TableFont \begin{tabular}{#1} #4
  \end{tabular}\end{center}\end{table}}

\newcommand{\MakeTablee}[5]{\begin{table}[htb]\TTabCap{#2}{#3}{#4}
  \begin{center} \TableFont \begin{tabular}{#1} #5
  \end{tabular}\end{center}\end{table}}

\newcommand{\MakeTableH}[4]{\begin{table}[H]\TabCap{#2}{#3}
  \begin{center} \TableFont \begin{tabular}{#1} #4
  \end{tabular}\end{center}\end{table}}

\newcommand{\MakeTableHH}[4]{\begin{table}[H]\TabCapp{#2}{#3}
  \begin{center} \TableFont \begin{tabular}{#1} #4
  \end{tabular}\end{center}\end{table}}

\newfont{\bb}{ptmbi8t at 12pt}
\newfont{\bbb}{cmbxti10}
\newfont{\bbbb}{cmbxti10 at 9pt}
\newcommand{\uprule}{\rule{0pt}{2.5ex}}
\newcommand{\douprule}{\rule[-2ex]{0pt}{4.5ex}}
\newcommand{\dorule}{\rule[-2ex]{0pt}{2ex}}
\def\thefootnote{\fnsymbol{footnote}}
\begin{Titlepage}
\Title{The OGLE Collection of Variable Stars.\\
Over 1.3 Million Human-Verified Eclipsing and Ellipsoidal Binaries\\in the Milky Way\footnote{Based on observations obtained with the 1.3-m Warsaw telescope at the Las Campanas Observatory of the Carnegie Institution for Science.}}
\Author{I.~~S~o~s~z~y~{\'n}~s~k~i$^1$,~~
M.~~G~{\l}~o~w~a~c~k~i$^1$,~~
M.\,A.~~U~r~b~a~n~o~w~i~c~z$^1$,~~
A.~~U~d~a~l~s~k~i$^1$,\\
M.\,K.~~S~z~y~m~a~{\'n}~s~k~i$^1$,~~
P.~~P~i~e~t~r~u~k~o~w~i~c~z$^1$,~~
J.~~S~k~o~w~r~o~n$^1$,~~
D.\,M.~~S~k~o~w~r~o~n$^1$,\\
R.~~P~o~l~e~s~k~i$^1$,~~
P.~~M~r~{\'o}~z$^1$,~~
P.~~I~w~a~n~e~k$^1$,~~
M.~~W~r~o~n~a$^{2,1}$,~~
S.~~K~o~z~{\l}~o~w~s~k~i$^1$,\\
K.~~U~l~a~c~z~y~k$^{3,1}$,~~
K.~~R~y~b~i~c~k~i$^{1}$,~~
M.~~G~r~o~m~a~d~z~k~i$^1$,~~
M.~~M~r~{\'o}~z$^1$,\\
J.~~B~o~r~o~w~i~c~z$^1$,~~
K.~~R~u~t~k~o~w~s~k~i$^1$,~~
M.~~C~e~l~i~{\'n}~s~k~i$^1$,~~
K.~~K~o~m~a~r$^1$,\\
P.~~M~a~t~e~r~a$^1$,~~
and~~J.~~W~a~l~i~c~k~a$^1$
}
{$^1$Astronomical Observatory, University of Warsaw, Al.~Ujazdowskie~4, 00-478~Warszawa, Poland\\
$^2$Villanova University, Department of Astrophysics and Planetary Sciences, 800 Lancaster Ave., Villanova, PA 19085, USA\\
$^3$Department of Physics, University of Warwick, Gibbet Hill Road, Coventry, CV4~7AL,~UK
}
\Received{\today}
\end{Titlepage}

\Abstract{We present a new release of the OGLE collection of eclipsing and ellipsoidal binary systems in the Galactic bulge and disk. The catalog contains 1\,330\,024 objects, including 1\,280\,765 eclipsing binaries and 49\,259 ellipsoidal variables, nearly three times as many systems as the previous editions of the OGLE collection. The selection and classification process involved expert visual inspection of the light curves. The catalog incorporates all available OGLE-II, OGLE-III, and OGLE-IV photometry, extending the observational baseline to nearly 28 years for some objects. The orbital periods and other observational parameters of previously known systems have been updated. We present basic statistics of the sample and highlight selected rare and unusual systems.}{binaries: eclipsing -- Catalogs}

\Section{Introduction}
Eclipsing binary systems play a crucial role in many areas of modern astrophysics. When precise time-series photometry and spectroscopy are available, eclipsing binaries provide the most direct and accurate measurements of fundamental stellar parameters, including masses, radii, effective temperatures, and luminosities (Andersen 1991, Torres \etal 2010, Southworth 2012), often with accuracy better than 1\% (\eg Ratajczak \etal 2021, Suchomska \etal 2022). These measurements serve as the primary empirical benchmarks for testing theories of stellar structure and evolution across a wide range of stellar masses, ages, and chemical compositions. Detached eclipsing binaries serve as geometric distance indicators and have been used to measure distances to nearby galaxies with unprecedented accuracy, providing one of the cornerstones of the extragalactic distance scale (Paczy{\'n}ski 1997, Pietrzy{\'n}ski \etal 2019).

Large samples of eclipsing binaries enable investigations of binary star formation and evolution, including tidal interactions, mass transfer, angular momentum loss, common-envelope evolution, and orbital circularization (\eg Zahn 1977, Eggleton 2006). Measurements of apsidal motion in eccentric systems constrain the internal structure of stars (Claret and Gimenez 1993), while eclipse timing variations may reveal additional companions (Borkovits \etal 2016, Hajdu \etal 2019). Eclipsing systems containing pulsating stars, such as Cepheids or $\delta$~Sct stars, offer unique opportunities to determine dynamical masses of pulsators and test pulsation theory (Pietrzy{\'n}ski \etal 2010, Pilecki \etal 2018).

The last three decades have witnessed a revolution in the discovery of eclipsing binaries driven by wide-field photometric surveys. Projects such as MACHO (Alcock \etal 1997), EROS (Kim \etal 2014), ASAS (Pojma{\'n}ski 2002), Catalina (Drake \etal 2017), ATLAS (Heinze \etal 2018), ASAS-SN (Jayasinghe \etal 2019), ZTF (Chen \etal 2020), and Gaia (Mowlavi \etal 2023) have increased the number of known eclipsing binaries from thousands to millions. Among large-scale surveys, the Optical Gravitational Lensing Experiment (OGLE) has played a particularly important role by providing long-term, high-cadence photometric monitoring of the densest stellar regions in the southern sky. The OGLE observations span approximately three decades and include billions of precise photometric measurements in the Galactic bulge, Galactic disk, and the Magellanic Clouds.

The OGLE Collection of Variable Stars (OCVS) contains more than one million objects spanning a wide range of variability types. What distinguishes the OCVS from other large catalogs of variable stars is that the final classification of each object is performed by a human expert based on a visual inspection of its light curve. As a result, the OCVS has an exceptionally low level of contamination from other types of variable objects and serves as a reference catalog for other variability surveys.

Eclipsing and ellipsoidal variables constitute more than half of variable stars included in the OCVS. The latest edition of the OGLE Collection of Eclipsing Variables in the Magellanic Clouds (G{\l}owacki \etal 2024) contains over 75\,000 objects. For the Milky Way, the OGLE team has published a sample of approximately 11\,000 eclipsing variables in selected regions (in total about 7 square degrees) of the Galactic disk (Pietrukowicz \etal 2013), as well as a catalog of more than 450\,000 eclipsing and ellipsoidal binary systems in the central regions of the Galactic bulge (Soszy{\'n}ski \etal 2016).

The OGLE Collection of Eclipsing and Ellipsoidal Binary Systems has found a remarkably wide range of astrophysical applications. This extensive sample has led to the discovery of numerous unusual objects, including doubly eclipsing systems (Zasche \etal 2019), HW Vir stars (Schaffenroth \etal 2019), candidates for dormant black holes in ellipsoidal variables (Gomel \etal 2021), double-periodic variables (Rojas Garc{\'\i}a \etal 2021), heartbeat stars (Wrona \etal 2022a), and ultra-short-period contact binaries (Papageorgiou \etal 2023). The long-term OGLE photometry has been used to investigate eclipse timing variations in searches for potential stellar merger candidates (Pietrukowicz \etal 2017, Hong \etal 2022, 2024) and triple systems (Hajdu \etal 2019, 2022). The light curves of eclipsing binaries have also been analyzed for additional periodic signals caused by stellar pulsations ({\'A}d{\'a}m \etal 2023). Other authors have explored tidally excited oscillations in eccentric ellipsoidal binaries (Wrona \etal 2022b), the O'Connell effect (Kleftogiannis \etal 2025), and the reflection effect (Frank \etal 2026). Physical parameters have been determined for selected binary systems included in the OCVS (Hong \etal 2017, Priya \etal 2020, Suchomska \etal 2022, Wang \etal 2024). Finally, the OGLE Collection of Eclipsing Binaries has become an indispensable training and validating set for machine-learning classification algorithms (\eg Becker \etal 2020, B{\'o}di and Hajdu 2021, Molnar \etal 2022, Szklen{\'a}r \etal 2022, Monsalves \etal 2024, Parimucha \etal 2025, Ula{\textcommabelow s} 2026) and for assessing the completeness and contamination of variable-star catalogs produced by other large-scale time-domain surveys (\eg Heinze \etal 2018, Christy \etal 2023, Gavras \etal 2023).

In this paper, we present a new release of the OGLE Collection of Eclipsing and Ellipsoidal Variables in the Galactic bulge and disk. The catalog contains nearly three times as many binary systems as the previous editions of the OGLE collection (Pietrukowicz \etal 2013, Soszy{\'n}ski \etal 2016), covering a region of the sky more than an order of magnitude larger than before. For previously known variables, we provide additional observations from the fourth phase of the OGLE project (OGLE-IV), extending their light curves by up to 10 years. For all objects, we provide updated orbital periods and other observational parameters, and for some systems we also revise their classification. Our collection can be complemented by other OGLE catalogs of binary systems representing specific classes, such as a sample of more than 1000 dwarf novae (Mr{\'o}z \etal 2015), a collection of nearly 1000 heartbeat stars (Wrona \etal 2022a), a list of nearly 100 candidates for planetary transits (Mr{\'o}z \etal 2023), and the recently published catalog of 30 circumbinary disk occultation systems (KH 15D-like stars; Urbanowicz \etal 2026).

The remainder of this paper is organized as follows. Section~2 describes the OGLE observational datasets. Section~3 outlines the criteria used to select and classify binary systems. Section~4 describes the collection of eclipsing and ellipsoidal variables in the Galactic bulge and disk. In Section 5, we present the catalog statistics, highlight selected objects of particular interest, and compare our collection with the Gaia DR3 catalog of eclipsing-binary candidates. Finally, Section 6 summarizes our results.

\Section{Observations and Data Reduction}
The time-series photometric data used in this study were obtained during the OGLE-IV survey, which operated from March 2010 to February 2025, with a two-year interruption due to the COVID-19 pandemic (2020--2022). The OGLE observations were collected with the 1.3-m Warsaw Telescope at Las Campanas Observatory (Chile), operated by the Carnegie Institution for Science. The telescope is equipped with a mosaic camera comprising 32 CCD detectors, each with a format of $2048\times4102$~pixels, providing a total field of view of approximately 1.4~square degrees and a pixel scale of 0.26~arcsec. OGLE observations are obtained predominantly through the Cousins {\it I}-band filter, supplemented by less frequent observations in the Johnson {\it V} band. The photometric reductions were performed with the Difference Image Analysis pipeline (Alard and Lupton 1998, Wo{\'z}niak 2000). Details of the OGLE instrumentation, data reduction procedures, and astrometric calibrations can be found in Udalski \etal (2015).

The OGLE-IV observations of the Milky Way were conducted in two modes. The standard OGLE fields, covering 182~square degrees in the central regions of the Galactic bulge, were observed with exposure times of 100~s in the {\it I} band and 150~s in the {\it V} band. The saturation limit in this region is approximately $I=13$~mag, while the faint limit depends on the degree of crowding and observing conditions, typically reaching $I=21$~mag. A much larger area of approximately 3000~square degrees in the Galactic disk and the outer regions of the bulge was covered by a subproject of the OGLE-IV survey known as the Galaxy Variability Survey (GVS). Observations in this area were obtained with exposure times of 25~s and 30~s in the {\it I} and {\it V} bands, respectively, yielding useful light curves for stars with {\it I}-band magnitudes in the range 11--19~mag.

The number of observations and the temporal coverage of the OGLE-IV light curves vary considerably between fields, ranging from several dozen to about $18\,000$ measurements, with time spans of 2 to 15~years. For some stars in our catalog, we also provide observations obtained during the OGLE-II (1997--2000) and OGLE-III (2001--2009) phases of the project. Combining these data with the OGLE-IV observations can extend the temporal baseline of the light curves to up to 28~years. We note that merging light curves from different phases of the OGLE project may require an adjustment of the zero point, which is best performed after converting the photometry from magnitudes to fluxes.

\Section{Selection and Classification od Binary Systems}
The OGLE Collection of Eclipsing and Ellipsoidal Binary Systems contains objects discovered using a variety of methods. Some were identified while searching for variable stars of other types, such as Cepheids (\eg Udalski \etal 2018), RR Lyr stars (\eg Soszy{\'n}ski \etal 2019), $\delta$ Sct stars (\eg Soszy{\'n}ski \etal 2021), or Blue Large-Amplitude Pulsators (BLAPs; \eg Borowicz \etal 2025). We also developed a semiautomated method for detecting eclipsing and ellipsoidal variable candidates based on light-curve template fitting (Soszy{\'n}ski \etal 2016). Additionally, a substantial fraction of the variables were added to the catalog through cross-identification of the OGLE photometric database with existing catalogs of eclipsing variables, in particular the International Variable Star Index (VSX; Watson \etal 2006) and the Gaia DR3 catalog of eclipsing variable candidates (Mowlavi \etal 2023). All these methods were used in parallel over a period of several years. Moreover, some binary systems were independently identified using more than one method. Therefore, it is currently not possible to determine how many stars were identified using each of these approaches.

Regardless of the method used for the initial selection of candidates for binary systems, the final stage of the procedure involved a visual inspection of their OGLE-IV {\it I}-band light curves. Variable stars that were classified by a human expert as eclipsing or ellipsoidal systems then underwent a refinement of their preliminary orbital periods. For this purpose, we used the programs {\sc Tatry} (Schwarzenberg-Czerny 1996), {\sc Fnpeaks}\footnote{http://helas.astro.uni.wroc.pl/deliverables.php?lang=en\&active=fnpeaks}, and {\sc IHSnpeaks}\footnote{https://github.com/kkrutkowski/ihsnpeaks} (Rutkowski and Poleski 2026), as well as the string-length algorithm (Dworetsky 1983). This procedure produced a list of possible orbital periods for each star, from which we selected the best one by fitting an eclipsing-binary light-curve model. For some light curves, particularly those affected by additional variability and therefore more noisy, the periods were adjusted manually.

We emphasize that we made every effort to ensure that our catalog contains the correct orbital periods. Nevertheless, we cannot rule out that the periods of some individual systems are incorrect. This may be particularly relevant for detached eclipsing systems for which only a few observations during eclipses were obtained. In such cases, the true orbital period may be twice as long or twice as short as the period listed in the catalog. An incorrect period can also result from a daily alias caused by the combination of the daily sampling of the OGLE data and the orbital frequency or one of its harmonics. In the future, additional OGLE observations should help verify and, where necessary, revise the periods given in our collection.

In the next stage, the sample of binary systems was divided into contact eclipsing binaries, non-contact eclipsing binaries (comprising detached and semi-detached systems), and ellipsoidal variables. We also identified 25 eclipsing cataclysmic variables. We note that our classification of binary systems into subclasses is based exclusively on the available OGLE-IV photometry and should be confirmed using other data, in particular spectroscopic observations. The uncertainty of our classification is greatest for the faintest stars, owing to the noise in their light curves.

The preliminary classification was performed using an automated method. For this purpose, each light curve was fitted with a phenomenological model consisting of a second-order Fourier series to describe the out-of-eclipse variability and up to two eclipses modelled using the profile introduced by Mikul{\'a}{\v{s}}ek (2015). The model was fitted using robust least squares with iterative outlier clipping. The resulting fit provides the Fourier coefficients and, for each eclipse, its phase, depth, width, and shape parameter, along with their uncertainties and the residual scatter. Based on these fitted parameters, additional quantities characterizing the light curve were calculated, including the total amplitude, the fraction of the orbital cycle spent in eclipse, the fraction of the variability attributable to the out-of-eclipse component, and the eclipse depth relative to its detection limit. The similarity between the two minima was assessed using the secondary-to-primary eclipse depth ratio and the ratio of the first two Fourier coefficients of the model. Based on these quantities and the orbital period, each star was assigned to one of three classes: contact, non-contact, or ellipsoidal binary. A probability of correct classification was also assigned to each star.

The classifier was calibrated against the OGLE-IV collection of eclipsing and ellipsoidal binaries toward the Galactic bulge (Soszy{\'n}ski \etal 2016). Among the 437\,885 stars in common, the classifier reproduces the OGLE classifications in 83.7\% of cases, with recall values of 0.66, 0.89, and 0.80 and for the contact, non-contact and ellipsoidal variables, respectively. For stars with a classification probability of at least 0.9, which account for 54.8\% of the sample, the agreement with the OGLE classification rises to 98.4\%. The classifier was used for the initial classification, which was subsequently verified through visual inspection of the light curves.

The photometry of the binary systems included in the previous editions of the OGLE collection (Pietrukowicz \etal 2013; Soszy{\'n}ski \etal 2016) has been updated. The light curves now include all available OGLE observations in the {\it I} and {\it V} bands obtained through February 2025. We note that the photometry of stars in the central regions of the Galactic bulge was re-reduced, and therefore the light curves of individual objects may differ slightly from those published in the catalog of Soszy{\'n}ski \etal (2016).

Taking advantage of the longer time baseline provided by the OGLE-IV data, we refined the orbital periods and other observational parameters (magnitudes, amplitudes, etc.) of previously known eclipsing and ellipsoidal variables. Following a visual inspection of the updated light curves, we decided to remove about 60 previously published objects from the OGLE Collection of Eclipsing and Ellipsoidal Variables, as we found them to be variable stars of other types, such as pulsating stars, spotted variables, or other (usually unidentified) variables. For 177 stars in the catalog, we revised their classifications, most commonly reclassifying ellipsoidal variables as eclipsing variables.

\Section{The OGLE Collection of Eclipsing and Ellipsoidal Binaries}

The final version of the OGLE Collection of Binary Systems in the Milky Way contains 1\,330\,024 objects, of which 1\,280\,765 ($\sim96.3$\%) are eclipsing binaries, while 49\,259 ($\sim3.7$\%) have been classified as ellipsoidal variables. About 55\% of the systems are observed toward the Galactic bulge region, while the remaining 45\% are located in the Galactic disk.

The complete list of all variables, together with their basic observational parameters (coordinates, orbital periods, {\it I}- and {\it V}-band magnitudes, primary and secondary eclipse depths, and the epoch of primary eclipse minimum), as well as all available photometric data obtained during the OGLE-II (1997--2000), OGLE-III (2001--2009), and OGLE-IV (2010--2025) project, can be downloaded from the OGLE Internet Archive:
\begin{center}
{\it https://www.astrouw.edu.pl/ogle/ogle4/OCVS/blg/ecl/}\\
{\it https://www.astrouw.edu.pl/ogle/ogle4/OCVS/gd/ecl/}\\
\end{center}

In the future, the collection will also be available through the OCVS WWW interface:
\begin{center}
{\it https://ogle.astrouw.edu.pl $\rightarrow$ OGLE Collection of Variable Stars}\\
\end{center}

\begin{figure}[p]
\includegraphics[width=12.6cm]{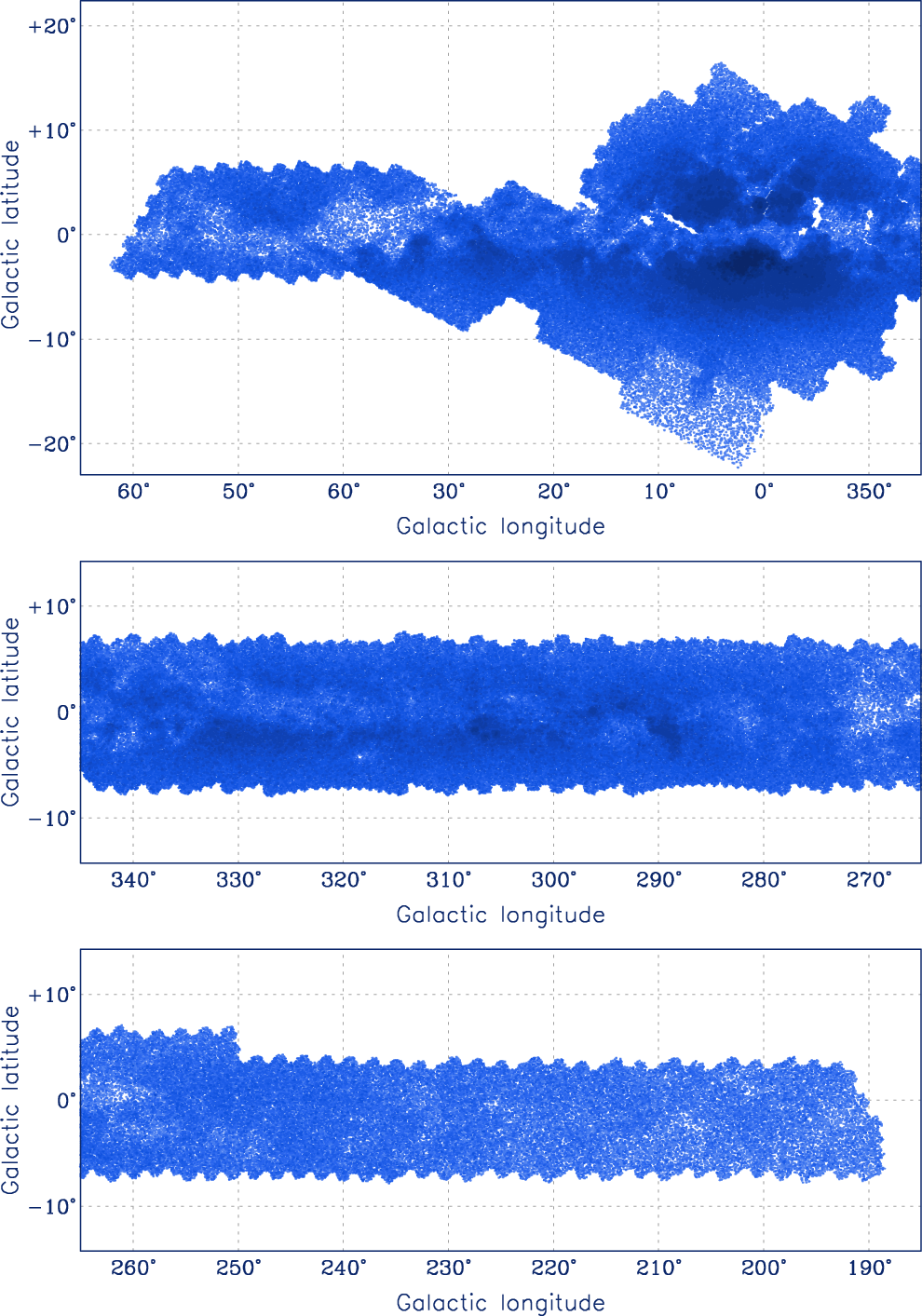}
\vspace*{2pt}
\FigCap{On-sky distribution of eclipsing and ellipsoidal binary systems in the Galactic bulge and disk. Darker colors indicate regions of higher stellar density.}
\end{figure}

Fig.~1 shows the sky distribution of more than 1.3 million eclipsing and ellipsoidal variables. The highest density, exceeding 10\,000 binary systems per square degree, is found near the Galactic center, in the so-called Baade's Window $(l, b)=(2^\circ,-3^\circ)$. On the opposite side of the sky, near the Galactic anticenter, we detected only about 70 eclipsing systems per square degree.

The variations in the sky distribution of stars visible in Fig.~1 are primarily caused by differences in actual stellar density in different regions of the Galaxy and differences in interstellar extinction along different lines of sight. This map also reflects the efficiency of our detection of eclipsing variables, which depends on the depth of the photometry (standard OGLE fields in the central regions of the bulge \vs the shallower GVS fields) and the temporal coverage of the light curves in individual OGLE fields. For example, the upper panel of Fig.~1 shows a sharp drop in the number of eclipsing and ellipsoidal stars in the southernmost part of the bulge ($b<-15^\circ$), which is caused by the small number of observations ($<50$) obtained so far by the OGLE survey in this region of the sky. In the future, we plan to update the OGLE Collection of Eclipsing Systems in this area using newly acquired observations.

The completeness of our catalog was estimated using stars located in the overlapping regions of adjacent OGLE-IV fields. Binary systems observed in two fields had two independent opportunities for detection. We note that the final version of the collection includes only one light curve per star, usually the one with the larger number of measurements. We verified {\it a posteriori} that 70\,095 binary systems in our sample were observed in two OGLE-IV fields. For 45\,561 of these systems, we independently identified both counterparts, whereas for the remaining 24\,534 systems, only one counterpart was identified during the selection process. This yields a formal catalog completeness of approximately 79\%. Naturally, this value applies only to the sample of eclipsing and ellipsoidal variables that could potentially be detected using OGLE photometric data, \ie stars whose brightness falls within the magnitude range covered by OGLE (see Section~2) and located in the regions of the Galactic bulge and disk monitored by the survey. We also note that a few percent of the area covered by the OGLE-IV fields is not observed because of technical gaps between the CCD chips of the OGLE-IV mosaic camera.

\begin{figure}[t]
\includegraphics[width=12.6cm]{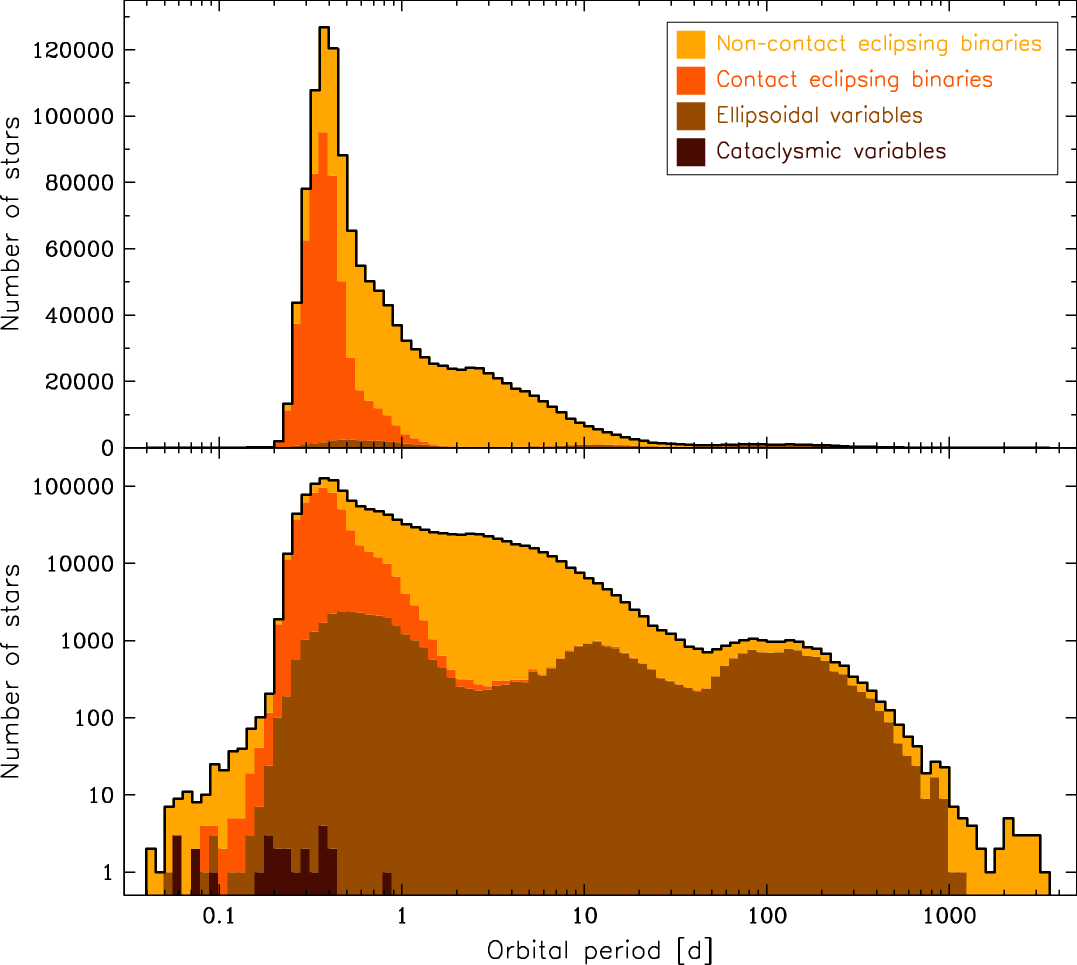}
\vspace*{2pt}
\FigCap{Distribution of orbital periods of eclipsing and ellipsoidal binary systems in the Milky Way. The yellow, red, light brown, and dark brown histograms represent non-contact eclipsing binaries, contact eclipsing binaries, ellipsoidal variables, and cataclysmic variables, respectively. The number of systems is shown on a linear scale in the upper panel and a logarithmic scale in the lower panel.}
\end{figure}

\Section{Discussion}

\Subsection{Catalog Statistics}
The OGLE Collection of Eclipsing and Ellipsoidal Variables in the Milky Way provides a valuable resource both for identifying rare and unusual binary systems and for conducting statistical studies of binary stars. Fig.~2 shows the distribution of orbital periods of the binary systems included in our catalog, plotted on linear (upper panel) and logarithmic (lower panel) scales. Our collection covers a wide range of orbital periods, from approximately one hour to more than 8~years. As shown in Figure 2, the distribution peaks sharply at around 0.4~d, with most systems in this peak classified as contact binaries (W~UMa-type stars). Two additional local maxima are visible in the histogram: one at orbital periods between 2 and 3~d and another at periods of around 100 days. The former is dominated by detached binaries containing main-sequence stars, while the latter consists primarily of systems containing red giant stars. A similar distribution of orbital periods has also been observed for binary systems in the Magellanic Clouds (G{\l}owacki \etal 2024).

\begin{figure}[t]
\includegraphics[width=12.6cm]{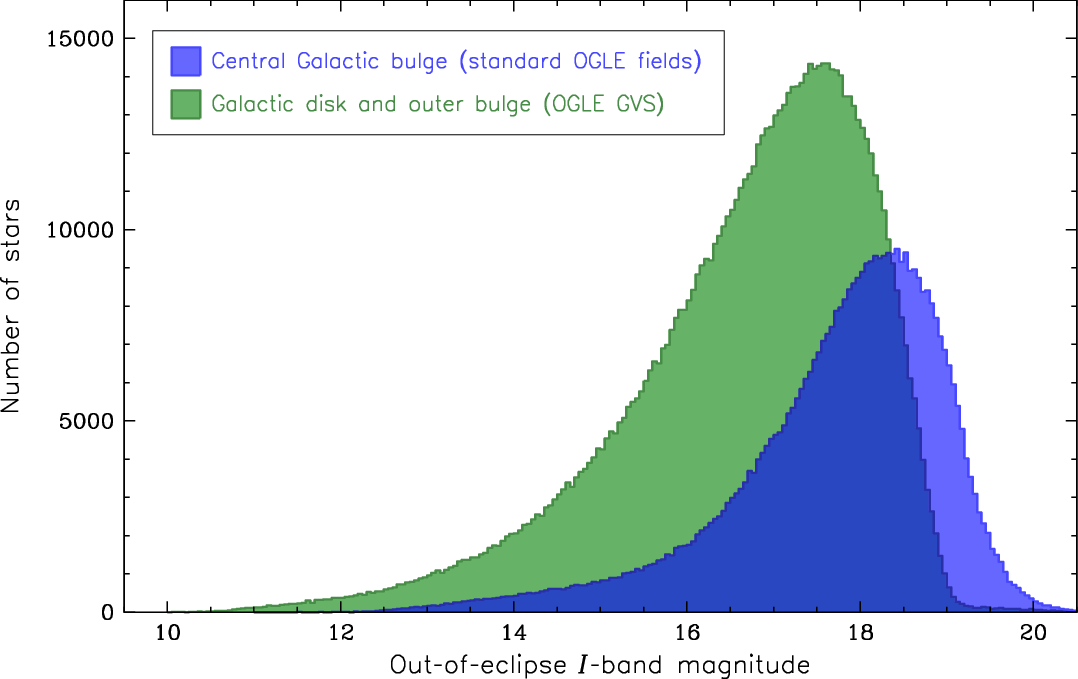}
\FigCap{Distribution of apparent out-of-eclipse {\it I}-band magnitudes of binary systems in the Milky Way. The blue and green histograms show systems observed toward the central Galactic bulge (standard OGLE fields) and the Galactic disk and outer bulge (GVS fields), respectively.}
\end{figure}

Fig.~3 presents the distribution of apparent out-of-eclipse {\it I}-band magnitudes of the binary stars in our sample. Different colors of the histograms distinguish between objects located in the standard OGLE fields covering the central Galactic bulge and those observed as part of the shallower GVS project, which covers the Galactic disk and the outer regions of the bulge. As expected, the different photometric depths of the two subsurveys of the OGLE project affect the completeness of our catalog. The magnitude distribution peaks at $I=18.4$~mag in the standard OGLE-IV fields and at $I=17.6$~mag in the GVS fields. Unlike in the central Galactic bulge, very few binary systems fainter than $I=19$~mag were detected in the Galactic disk.

\Subsection{Particularly Interesting Objects}
Among the 1.3 million eclipsing and ellipsoidal variables in our catalog, one can find systems representing virtually all types of binary stars, including rare and remarkable objects such as eclipsing cataclysmic variables (\eg Mr{\'o}z \etal 2015), doubly eclipsing systems (\eg Zasche \etal 2019), heartbeat stars (\eg Wrona \etal 2022ab), ultra-short-period binaries (\eg Papageorgiou \etal 2023), RS~CVn stars (\eg Iwanek \etal 2024), and even systems exhibiting planetary transits (\eg\linebreak Mr{\'o}z \etal 2023). Moreover, the long-term OGLE photometry enables studies of a wide range of time-dependent phenomena in binary systems, including changes in orbital periods (\eg Pietrukowicz \etal 2017), shifts in eclipse phases due to apsidal motion, variations in mean brightness, and changes in eclipse depths, including the disappearance and reappearance of eclipses (transient eclipsing binaries; Graczyk \etal 2011). All identified objects with unusual properties are flagged in the catalog remarks. However, this is not a comprehensive list of noteworthy objects, and we therefore encourage readers to explore the collection themselves.

Our collection includes eclipsing and ellipsoidal systems that exhibit additional variability not directly related to their binarity. More than 40 such systems contain pulsating stars, including Cepheids, $\delta$~Sct stars, and RR~Lyr-like stars, such as OGLE-BLG-RRLYR-02792 (Pietrzy{\'n}ski \etal 2012), designated OGLE-BLG-ECL-108266 in our catalog. These systems provide a powerful means of independently determining the masses and radii of pulsating stars, thereby enabling tests of stellar pulsation and evolution theories.

\begin{figure}[p]
\includegraphics[width=12.6cm]{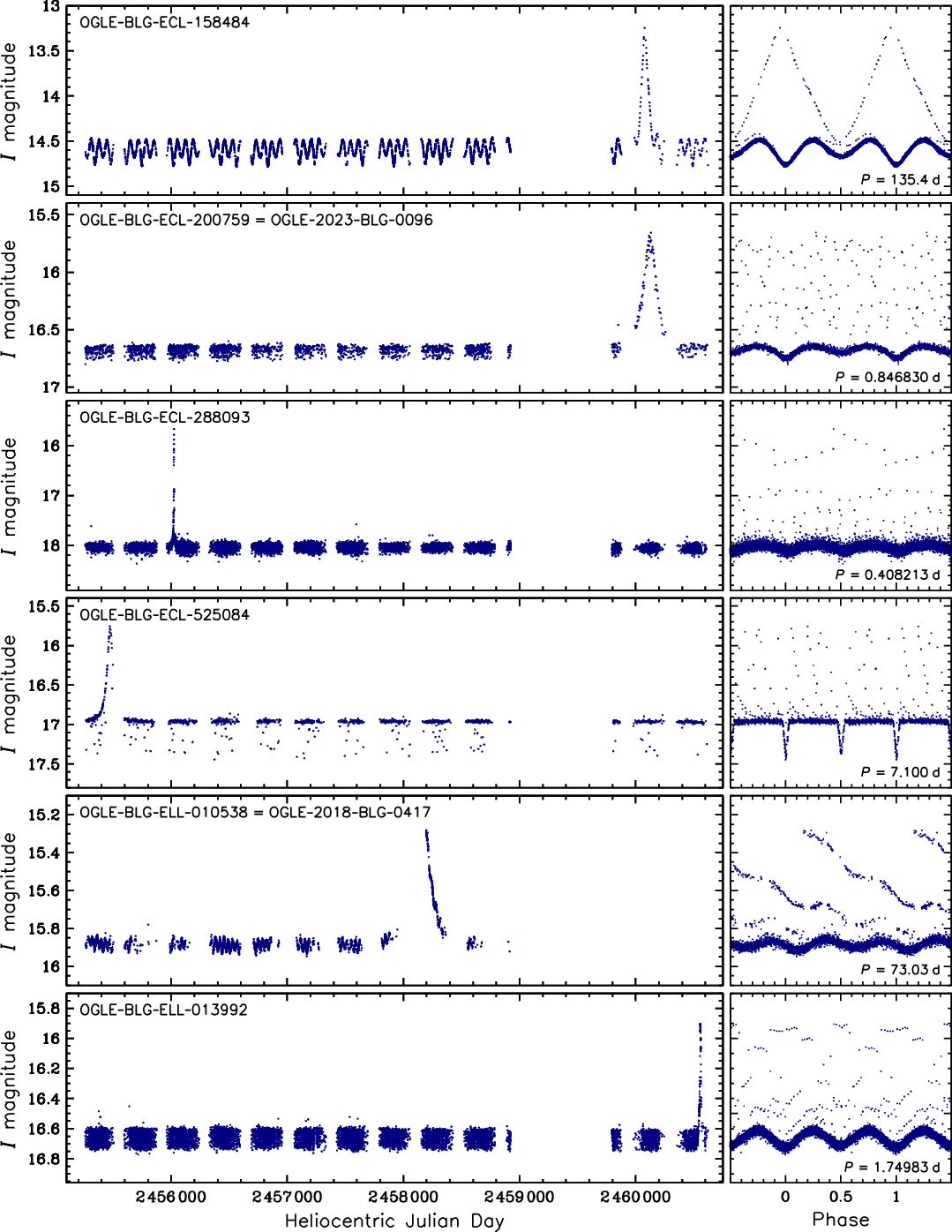}
\vspace*{3pt}
\FigCap{OGLE-IV {\it I}-band light curves of six eclipsing and ellipsoidal binaries that exhibited gravitational microlensing events. The left panels show the original light curves, while the right panels show the same light curves phased with the orbital periods.}
\end{figure}

During our search and classification process, we also detected more than a dozen eclipsing and ellipsoidal variables that exhibited gravitational microlensing events. The light curves of six representative objects of that type are presented in Fig.~4. Such events often go undetected in standard microlensing searches because stars exhibiting regular, large-amplitude variability are rejected at the initial selection stage. Indeed, only two of the six microlensing events shown in Fig.~4 were detected by the OGLE Early Warning System (Udalski \etal 2015). Microlensed variable stars may be particularly valuable for breaking degeneracies in microlensing models (\eg Wyrzykowski \etal 2006), thereby enabling more accurate determinations of the physical properties of the lens.

\begin{figure}[t]
\includegraphics[width=12.6cm]{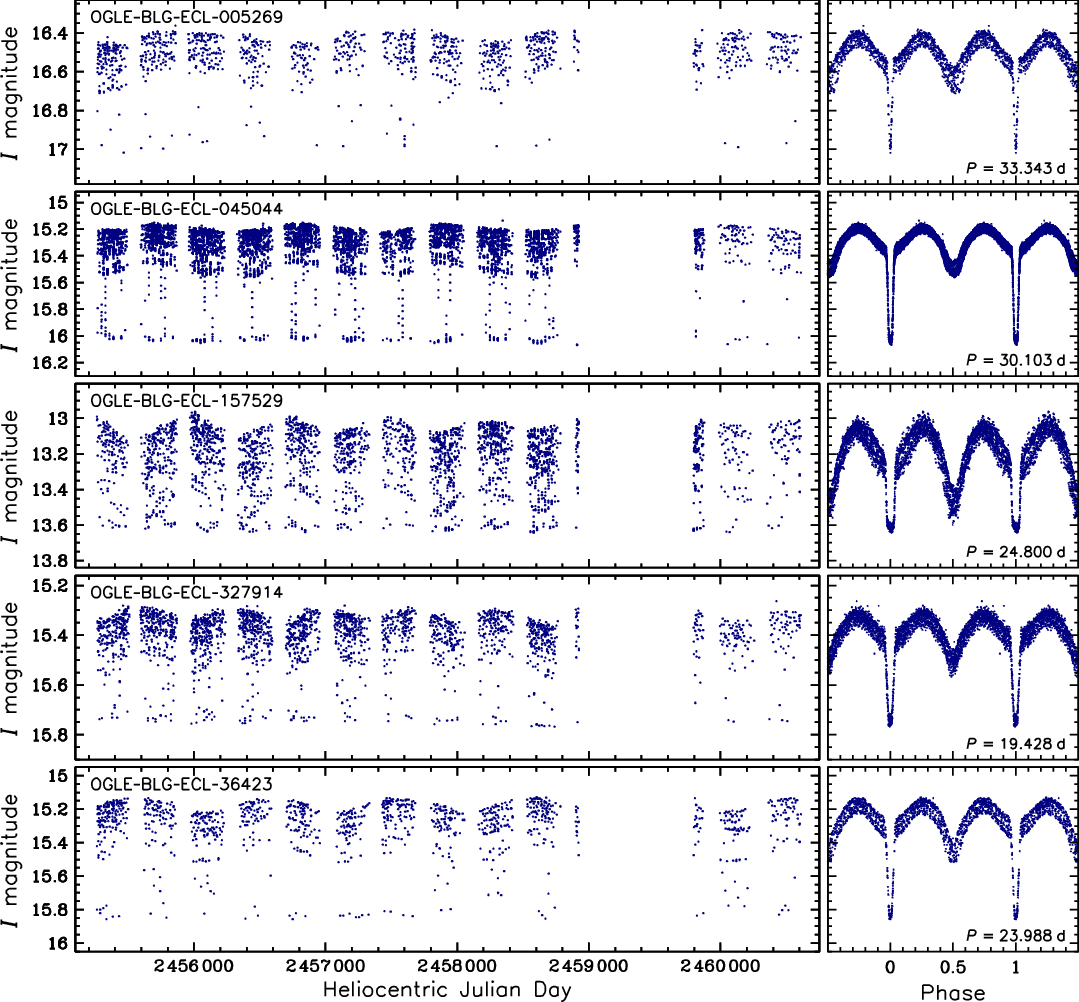}
\FigCap{OGLE-IV {\it I}-band light curves of five double-periodic variables. The left panels show the original light curves, revealing the long-term brightness variations, while the right panels show the same light curves phase-folded with the orbital periods.}
\end{figure}

In Fig.~5, we present the light curves of five of more than 30 identified double-periodic variables (\eg Rojas Garc{\'\i}a \etal 2021, Mennickent and Djura{\v{s}}evi{\'c} 2021). These objects belong to a rare class of interacting binary stars that exhibit two distinct periods of brightness variations: an orbital period associated with eclipses (or ellipsoidal modulation) and a long cycle, typically 30--40 times longer than the orbital period. The origin of the long cycle remains unclear. We hope that the newly discovered systems of this type, together with the extended OGLE light curves of the previously known double-periodic variables, will help shed light on the nature of these enigmatic objects.

Another relatively rare type of eclipsing variable comprises systems exhibiting the reflection effect, in which a hot star irradiates a cool companion. Examples include HW~Vir stars, \ie post-common-envelope systems composed of a hot B-type subdwarf and a cool main-sequence star (Schaffenroth \etal 2019), as well as nascent binaries in which a B-type main-sequence star irradiates a cool pre-main-sequence companion (Moe and Di Stefano 2015). In Fig. 6, we present the light curves of nine eclipsing variables exhibiting exceptionally large amplitudes of the reflection effect, in most cases exceeding 1~mag and reaching 2~mag in some cases. Such enormous brightness variations indicate an extreme temperature contrast between the components.

\begin{figure}[t]
\includegraphics[width=12.6cm]{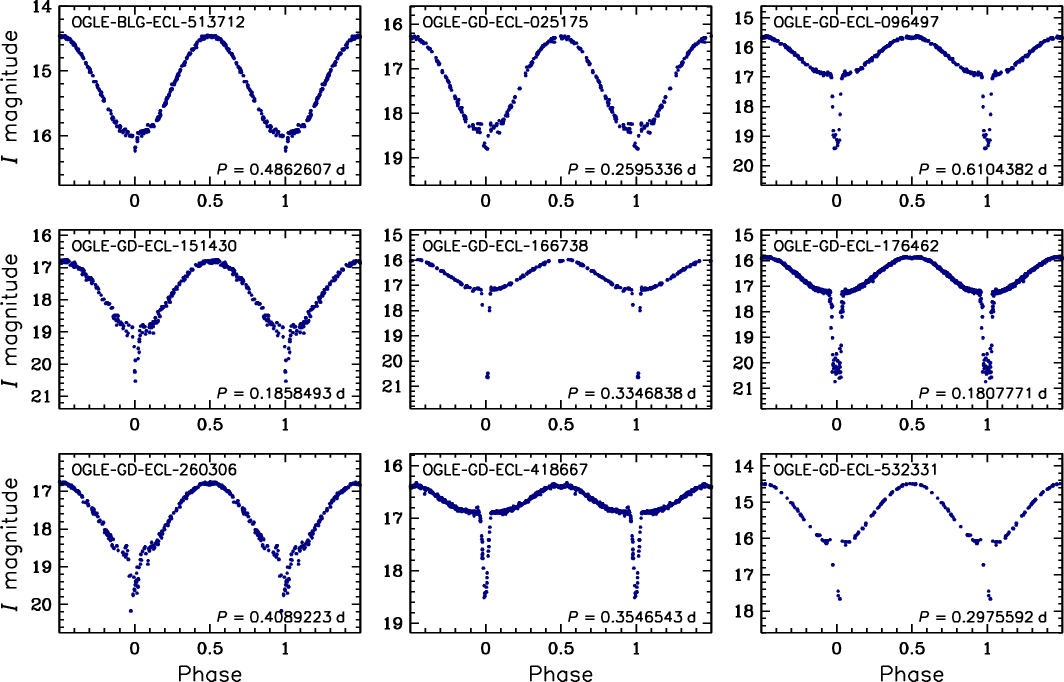}
\FigCap{OGLE-IV {\it I}-band light curves of nine eclipsing binary systems exhibiting an exceptionally large-amplitude reflection effect.}
\end{figure}

Despite the increasing use of machine-learning algorithms, manual selection and classification of variable stars offer unique advantages. Human researchers can recognize anomalies that automated systems may dismiss as noise or measurement errors. In Fig.~7, we present example light curves of eclipsing and ellipsoidal variables that caught our attention because of their unusual eclipse profiles or peculiar brightness variations outside eclipses. The atypical behavior of these stars may be attributable to circumstellar or circumbinary disks, starspots, tidally excited oscillations, or highly eccentric orbits. We encourage readers to explore our collection for similarly unusual objects.

\begin{figure}[t]
\includegraphics[width=12.6cm]{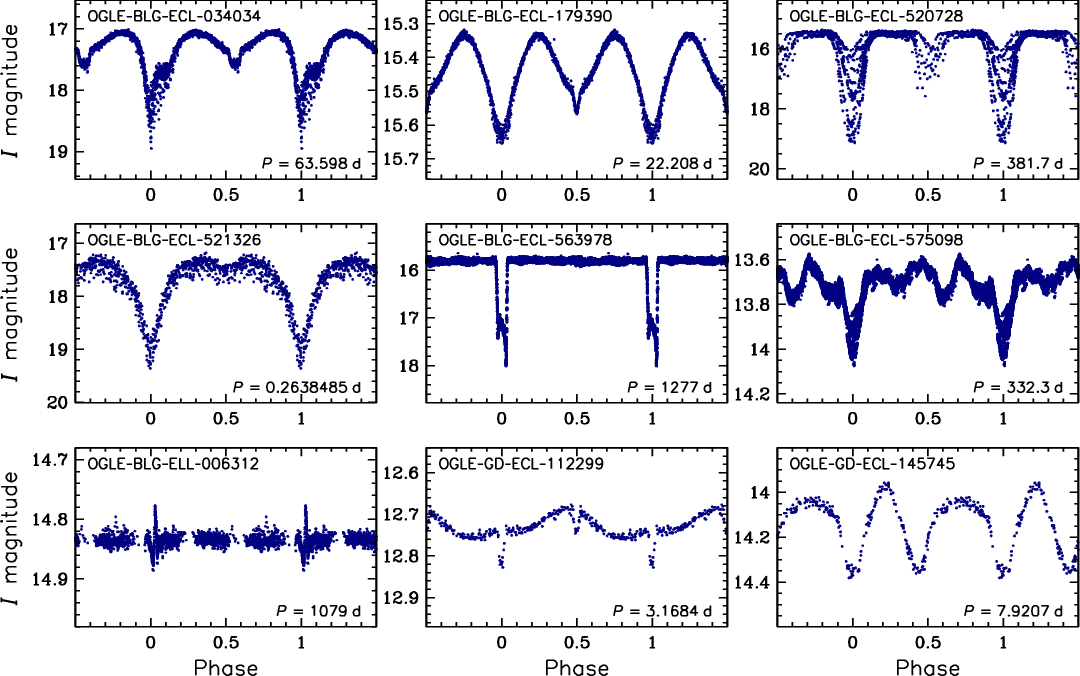}
\FigCap{OGLE-IV {\it I}-band light curves of nine eclipsing and ellipsoidal variables exhibiting unusual photometric behavior.}
\end{figure}

\Subsection{Comparison to the Gaia DR3 Catalog}
During the search for binary systems in the OGLE database, we examined all eclipsing candidates from the Gaia DR3 catalog (Mowlavi \etal 2023) located within the OGLE fields, allowing us to estimate the purity of the Gaia sample. The full Gaia DR3 catalog contains 2\,184\,477 candidates for eclipsing binaries distributed across the entire sky. Within the OGLE fields covering the Galactic bulge and disk, we identified almost exactly half of these candidates (1\,095\,756). However, the final version of our collection of eclipsing and ellipsoidal variables includes only 703\,237 objects listed in the Gaia DR3 catalog, corresponding to 64.2\% of the Gaia candidates located within the OGLE fields. We could not confirm the binary nature of the remaining 392\,519 candidates (35.8\%), although we cannot rule out the possibility that some are too faint or have too few measurements in the OGLE light curves for their variability type to be reliably determined.

A more reliable estimate of the purity of the Gaia DR3 sample can be obtained for the central regions of the Galactic bulge, which are covered by the standard OGLE-IV fields. In these regions, the OGLE photometry reaches a depth comparable to that of Gaia DR3 (down to $I\approx21$~mag), while the large number of measurements (from several hundred to over 18\,000 per star) makes the detection of eclipsing variability highly reliable. We found that 196\,089 Gaia DR3 eclipsing candidates are located within the 182 square degrees covered by the standard OGLE-IV fields, but only 93\,709 (47.8\%) of these stars are included in our collection. Among the remaining 102\,380 (52.2\%) Gaia candidates for eclipsing binaries, we identified several thousand variable stars of other types (\eg pulsating, spotted, and long-period variables), but most show no significant variability in the OGLE data. These results suggest that the purity of the Gaia DR3 sample decreases significantly toward the Galactic center, possibly owing to the higher stellar density in these regions.

Additionally, we compared the orbital periods derived from the Gaia DR3 and OGLE light curves. For 84\% of the 703,237 binary systems included in both the Gaia and OGLE catalogs, the derived orbital periods agree to within 1\%. This fraction increases to 89\% when period ratios of two or one-half are also considered. We also note that Mowlavi \etal (2023) compared the Gaia DR3 catalog of eclipsing-binary candidates with the OGLE catalogs covering the Magellanic Clouds (Pawlak \etal 2016) and the Galactic bulge (Soszy{\'n}ski \etal 2016) and estimated the completeness of their sample to be between 25\% and 50\%, depending on the region of the sky.

\Section{Conclusions}
We have presented the OGLE collection of more than 1.3~million eclipsing and ellipsoidal binary systems in the Milky Way. Compared with the previous editions of the OGLE catalog of binary systems in the Galactic disk and bulge (Pietrukowicz \etal 2013; Soszy{\'n}ski \etal 2016), the number of objects has increased by almost a factor of three. The total number of variable stars in the OCVS has now exceeded 2~million, making it the largest manually verified sample of variable stars ever assembled in astronomy.

Such a large collection of eclipsing and ellipsoidal binary systems offers a wealth of opportunities for astrophysical research, particularly when the long-term OGLE photometry is combined with the massive spectroscopic datasets provided by the Gaia mission. These data will enable the determination of fundamental stellar parameters on a large scale, the calibration of stellar evolution models, the study of binary populations, Galactic mapping, and the detection of black holes and planets. The nearly 30-year baseline of OGLE observations also allows us to investigate any kind of out-of-eclipse variability in binary systems, perform precise eclipse-timing analyses to identify and study multiple systems, and search for orbital period changes that may herald stellar mergers. Finally, the OCVS provides an excellent training and test dataset for machine-learning and deep-learning algorithms.

\Acknow{This work has been funded by the National Science Centre, Poland, grant no.~2022/45/B/ST9/00243. For the purpose of Open Access, the author has applied a CC-BY public copyright license to any Author Accepted Manuscript (AAM) version arising from this submission.}

\end{document}